      [1999/12/01 v1.4c Il Nuovo Cimento]
\documentclass{cimento}

\usepackage{graphicx} 
\usepackage{amssymb,amsmath}
\title{Standard model extensions for PV electron scattering, $g-2$, EDM: Overview}
\author{J.~Erler\from{ins:x}\from{ins:y}\thanks{erler@fisica.unam.mx}}
\instlist{\inst{ins:x} School of Natural Sciences, Institute for Advanced Study, \\
Einstein Drive, Princeton, NJ 08540, USA \\
\inst{ins:y} Permanent address: Departamento de F\'isica Te\'orica, Instituto de F\'isica, \\ 
Universidad Nacional Aut\'onoma de M\'exico, 04510 M\'exico D.F., M\'exico}
\PACSes{
\PACSit{12.60.Jv}{Supersymmetric models}
\PACSit{13.40.Em}{Electric and magnetic moments}
\PACSit{13.60.-r}{Photon and charged-lepton interactions with hadrons}
\PACSit{14.70.Pw}{Other gauge bosons}}
\begin{document}

\maketitle

\begin{abstract}
I review how various extensions of the Standard Model, in particular supersymmetry and extra
neutral gauge bosons, may affect low energy observables, including parity-violating electron scattering
and related observables, as well as electric and magnetic dipole moments.
\end{abstract}

\section{The Standard Model and Beyond}
\begin{figure}
\includegraphics[scale=0.51]{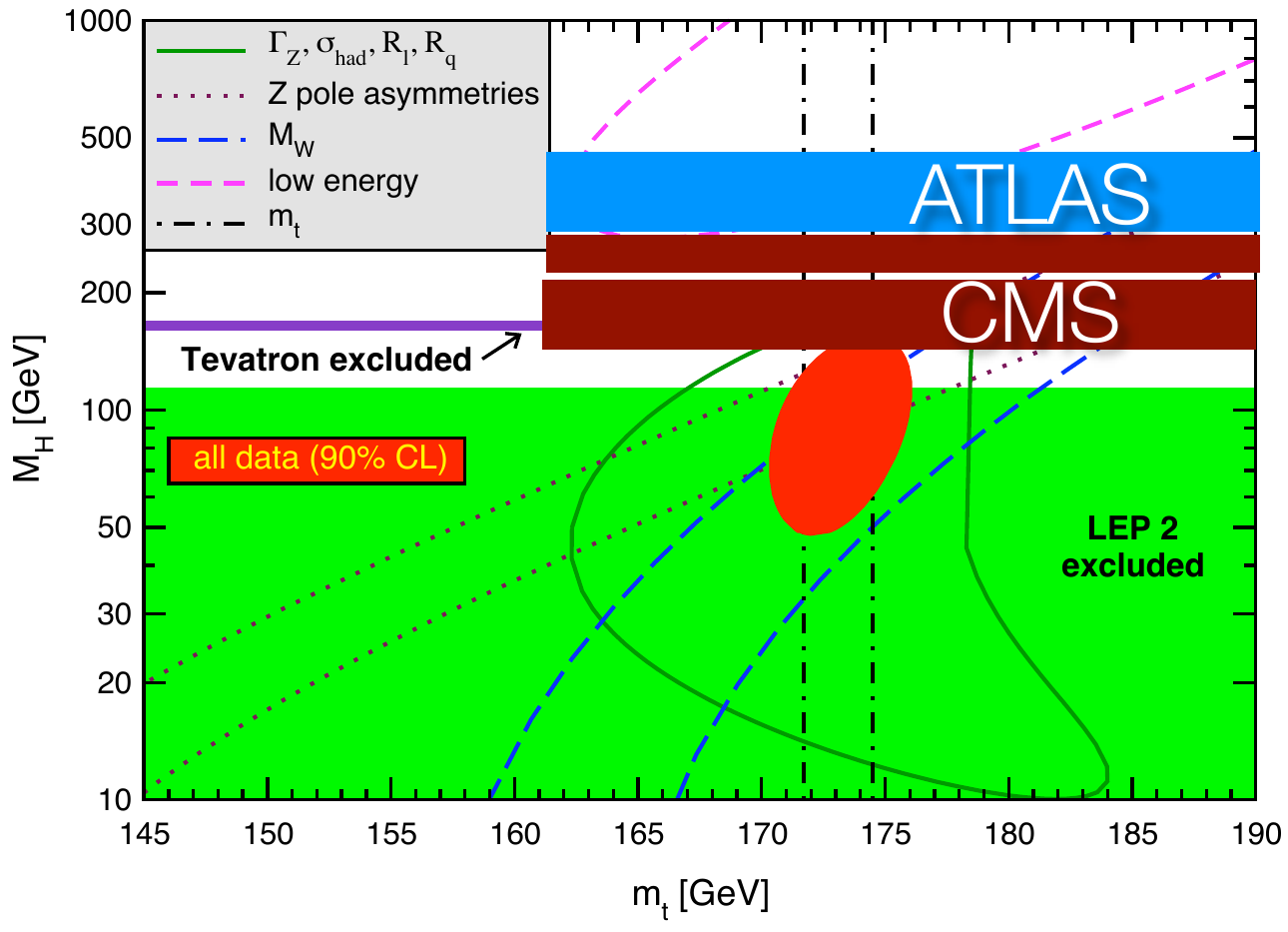}
\includegraphics[scale=0.51]{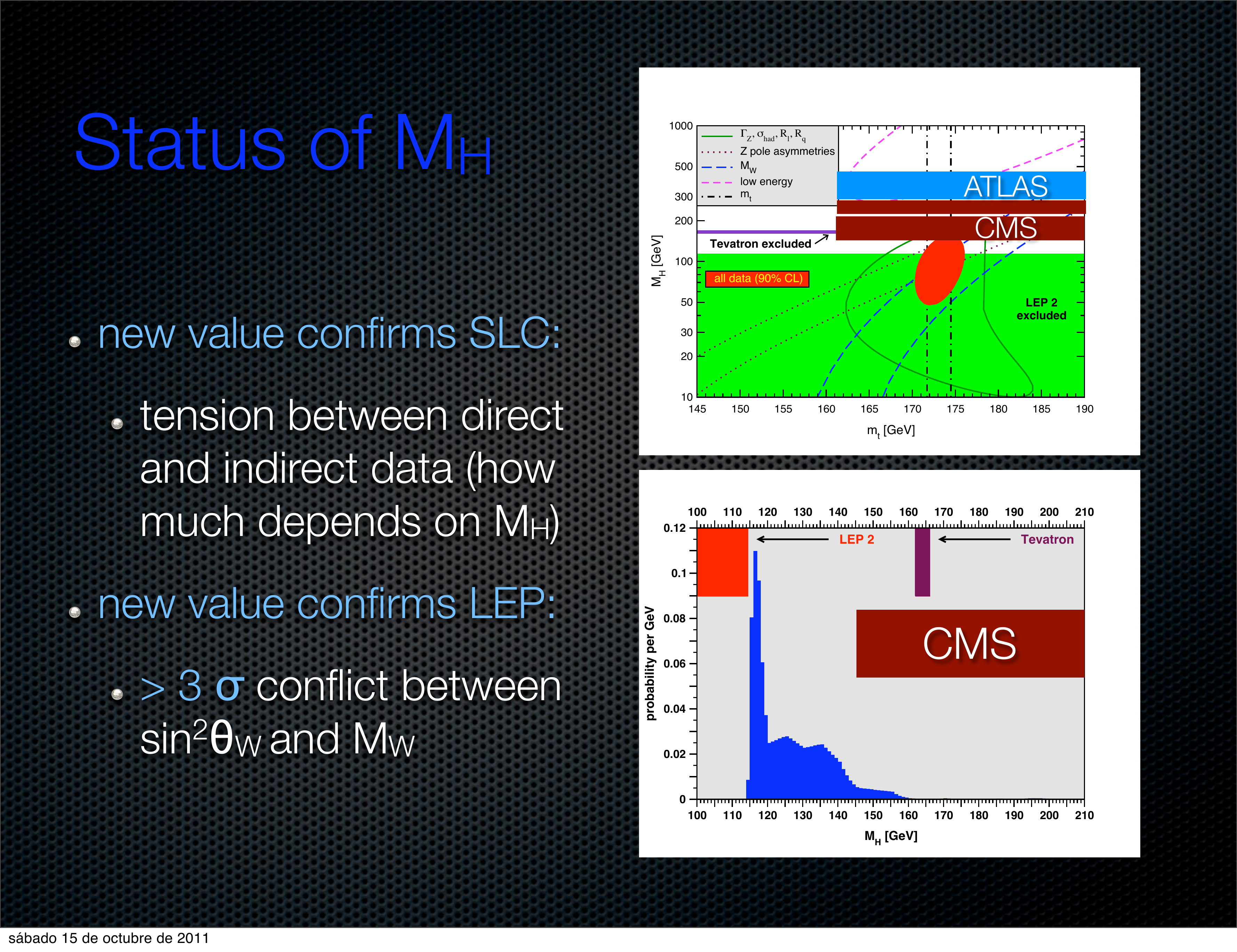}
\caption{Left: Contours of $M_H$ as a function of the top quark mass, $m_t$, for various EW data sets.
The red ellipse gives the 90\% CL allowed region by all precision data.
Also indicated are the 95\% exclusion intervals for $M_H$ from LEP~2, the Tevatron, ATLAS and CMS.
Right: Normalized probability distribution of $M_H$ based on the combination of EW precision data with
LEP~2 and Tevatron search results.  The CMS exclusion is seen to be fully consistent with this.}
\label{mh}
\end{figure}

The basic structure of the Standard Model (SM) has been well established through the discovery
of weak neutral currents, through parity violating deep inelastic scattering~\cite{Prescott:1978tm},
and through the discovery of the $W$ and $Z$ bosons.  
Moreover, the SM has been established as a spontaneously broken quantum field theory 
--- even though the Higgs boson has not been discovered, yet --- 
through the very high precision $Z$ factories LEP~1 and SLC.  
Finally, we are closing in on the Higgs boson.  
Figure~\ref{mh} shows the probability distribution of the Higgs mass, $M_H$, within the SM, 
based on all available electroweak (EW) precision data and search results from LEP~2 and the Tevatron.  
The remaining window, 115~GeV~$\lesssim M_H \lesssim 160$~GeV, is fully consistent with 
the negative search results at the LHC in almost the entire range, 145~GeV~$\lesssim M_H \lesssim 470$~GeV, 
and also with some $2~\sigma$ level excesses below it~\cite{delRe:PAVI2011}.  
The flip side of this spectacular success of the SM are depressed expectations for physics beyond it.  
There are some smaller deviations but nothing conclusive has arisen so far.  
Of course, neutrino masses and mixings indicated by neutrino oscillations are often seen as physics beyond the SM.  
But it should be cautioned that this class of effects can be fully described by non-renormalizable dimension~5 operators 
of the form $HH\bar{L}^c_iL_j$~\cite{Weinberg:1979sa} where $L_i$ are lepton doublets, 
and neutrino masses of roughly $10^{-5}$ to $10^{-1}$~eV had been anticipated beforehand~\cite{Weinberg:1979sa}.
In light of all this, why should one move beyond the SM?  

There is irrefutable evidence for dark matter, dark energy, and the excess of baryons over anti-baryons in the 
observable universe, which all need explaining and most likely the introduction of new degrees of freedom.
The hierarchy between the EW and Planck scales and the cosmological constant problem
present themselves as serious theoretical mysteries.
One may also point to the arbitrariness of the SM gauge group, $SU(3)_C \times SU(2)_L \times U(1)_Y$, 
the way it is represented in the particle spectrum and to the unexplained values (and hierarchies) of the SM
parameters.  There are even tantalizing hints at a possible unification structure (the $E_6$ gauge group
or a subgroup) and at gauge coupling unification in the minimal supersymmetric SM (MSSM).  

In any case, the overriding goal must be to uncover the {\em principles\/} underlying the SM.
The periodic table of the elements, for example, was a breakthrough, as it provided a better 
understanding of what were then just random facts of chemistry, and moreover produced 
successful predictions of missing entries (this is where we are today in elementary particle physics). 
Even more importantly it served Niels Bohr to argue in favor of the correspondence principle.  
Incidentally, constructed long before the neutron was discovered it was used to hypothesize it, and thus 
to add a new dimension to it, which is vaguely reminiscent of present day speculations about supersymmetry (SUSY).

Before discussing some new physics examples in more detail, this may be a good place to 
quote model-independently the energy scales, $\Lambda_{\rm new}$ (see Table~\ref{tab:sensitivities}),
which are typically probed by various kinds of observables.  In the electron scattering sector one can write,
\begin{equation}
   \Lambda_{\rm new} \simeq {1\over \sqrt{\sqrt{2}\, G_F \Delta Q_W}}
   = {246.22 \mbox{ GeV} \over \sqrt{\Delta Q_W}},
\end{equation}
where $G_F$ is the Fermi constant and $\Delta Q_W$ is the combined experimental and theoretical 
uncertainty in the (generalized) weak charges measured in polarization asymmetries.

Dimension~5 electromagnetic dipole moment operators are defined with help of the effective Lagrangian,
${\cal L}= 1/2 (D \bar\psi \sigma_{\mu\nu} P_R \psi + {\rm H.c.}) F^{\mu\nu}$,
where the anomalous magnetic moments (MDMs), $a$, are given in terms of $\Re e\, D = e a/(2 m)$, and
electric dipole moments (EDMs) can be defined by $d = \Im m\, D$.  One then has,
\begin{equation}
\Lambda_{\rm new} = {m_\mu\over \sqrt{\Delta a_\mu}}  \hspace{10pt} \mbox{ and }  \hspace{10pt}
\Lambda_{\rm new} = \sqrt{e m_e\over 2 \Delta d_e},
\end{equation}
for the muon MDM and the electron EDM, respectively.
These are related to similarly defined flavor transition moments as searched for in $\mu \to e$ conversion experiments. 

\begin{table}
  \caption{Typical sensitivities of past, current and future experiments to new physics scales, $\Lambda_{\rm new}$.
  The $C_{1j}$ are the coefficients of model-independent four-Fermi operators with vector couplings to quarks and
  axial-vector couplings to leptons, with the reverse being true of the $C_{2j}$.}
  \label{tab:sensitivities}
  \begin{tabular}{lrrr}
  \hline
  combination of couplings & $\Lambda_{\rm new}$ & experiment & laboratory \\
    \hline
      $2\, C_{2e}\equiv Q_W^e$ & 3.4 TeV & E158~\cite{Anthony:2005pm}& SLAC  \\
      $2\, C_{2e}\equiv Q_W^e$ & 7.5 TeV & MOLLER~\cite{Mammei:PAVI2011}& JLab \\
      $2\, (2\, C_{1u} + C_{1d})\equiv Q_W^p$  & 4.6 TeV & Qweak~\cite{Myers:PAVI2011} & JLab \\
      $2\, (2\, C_{1u} + C_{1d})\equiv Q_W^p$ & 6.3 TeV & P2~\cite{Maas:PAVI2011} & Mainz \\
      $2\, (2\, C_{1u} - C_{1d}) + 1.68\, (2\, C_{2u} - C_{2d})$ & 2.5 TeV & PVDIS~\cite{Zheng:PAVI2011} \& SoLID~\cite{Reimer:PAVI2011} & JLab \\
    \hline
      $a_\mu$  & 3.8 TeV & E821~\cite{Bennett:2006fi} & BNL \\
      $d_e$  & 83 TeV & $^{205}$Tl EDM~\cite{Regan:2002ta} & Berkeley \\
      $\mu \to e$  & 300 TeV & SINDRUM~II (Au)~\cite{Bertl:2006up} & PSI \\
    \hline
  \end{tabular}
\end{table}

\section{Illustrative example: supersymmetric extensions}
Theoretically, SUSY plays a central role in particle physics,
which is in part because it is the uniquely possible extension of the Poincar\'e group.
SUSY is also the only way to couple massless spin~3/2 particles
in much the same sense as massless spin~1 and spin~2 particles 
need gauge and local Lorentz symmetries, respectively. 
Finally, SUSY provides an elegant solution to the hierarchy problem, 
a property which may also be useful to stabilize possible large extra dimensions.
From the observational point of view one may point
to the fairly solid prediction for a light Higgs exactly as allowed by Fig.~\ref{mh},
to natural radiative EW symmetry breaking for realistic values of $m_t$, and
to the fact that the lightest supersymmetric particle is a viable dark matter candidate (if stable).

However, at least in its minimal (MSSM) version SUSY has some problems,
among the most pressing ones the so-called $\mu$-problem~\cite{Kim:1983dt}
(a remnant of the hierarchy problem) and the fact that dimension~4 proton decay 
is not disallowed through an  accidental symmetry as in the SM, which has to be seen as a step backwards.
One needs to introduce 105 new free parameters, plus additional ones if an {\em ad hoc} 
$R$-symmetry is broken or if there are non-holomorphic SUSY breaking terms. 
It is also unknown how SUSY is broken and how the breaking might be mediated from
some hidden sector to our observable world. 
Finally, the non-observation of superpartners and extra Higgs particles is discouraging 
and gives rise to what is called the little hierarchy problem, {\em i.e.\/}, 
the reemergence of the need for some parameter tuning.
One may thus treat the MSSM as an important reference model when analyzing 
and interpreting experimental data,
but one may expect additional ingredients to solve its problems (such as extra gauge symmetries).

\section{$g_\mu - 2$}
The experimental value~\cite{Bennett:2006fi} of $a_\mu$ can be written as,
\begin{equation}
  a_\mu^{\rm exp} \equiv {g_\mu - 2\over 2} = {\alpha\over 2 \pi} + (4511.17 \pm 0.63) \times10^{-9},
\end{equation}
giving rise to a discrepancy with theory, 
$a_\mu^{\rm exp} - a_\mu^{\rm th} = (2.88 \pm 0.80)\times10^{-9}$ (or $3.6~\sigma$),
when a SM prediction based on data from $e^+ e^- \to$ hadrons is used for the 2-loop
vacuum polarization contribution of $\Delta a_\mu(e^+ e^-) = (69.23 \pm 0.42) \times10^{-9}$~\cite{Davier:2010nc}.
A smaller deviation, $a_\mu^{\rm exp} - a_\mu^{\rm th} = (1.96 \pm 0.83)\times 10^{-9}$ ($2.4~\sigma$), is seen
if data from $\tau \to \nu_\tau +$ hadrons are used (where possible) instead, 
$\Delta a_\mu(\tau) = (70.15 \pm 0.47)\times10^{-9}$~\cite{Davier:2010nc}.
The conflict between the two data sets is
$\Delta a_\mu(\tau) - \Delta a_\mu(e^+ e^-) = (0.91 \pm 0.50) \times10^{-9}$ or $1.8~\sigma$.
Averaging yields $\Delta a_\mu(\mbox{average}) = (69.61 \pm 0.36)\times10^{-9}$ and
a deviation, $a_\mu^{\rm exp} - a_\mu^{\rm th} = (2.50 \pm 0.77)\times10^{-9}$ (or $3.2~\sigma$).
Note, that this hadronic correction is correlated with the $Z$ pole value of $\alpha$ and the low-energy
weak mixing angle, $\sin^2\theta_W(0)$, and thus with $M_H$. 
This connection has also been discussed in the context of new physics~\cite{Passera:2008jk}.

The above results include an additional uncertainty from hadronic 3-loop light-by-light scattering diagrams
which contribute, $\Delta a_\mu(\gamma\times \gamma) = (1.05 \pm 0.26)\times10^{-9}$~\cite{Prades:2009tw}.
This is consistent with the 95\% CL upper bound, $\Delta a_\mu(\gamma\times \gamma) < 1.59 \times10^{-9}$, 
found in Ref.~\cite{Erler:2006vu}.

One may point out that if the three dominant errors from experiment ($6.3\times10^{-10}$),
hadronic vacuum polarization ($3.6\times10^{-10}$) and light-by-light scattering ($2.6\times10^{-10}$)
can be pushed below $3\times10^{-10}$, 
then a $5~\sigma$ discovery would be established (if the central value persists).  
The $\Delta a_\mu(\gamma\times \gamma)$ error is already there, but it is also the hardest to defend.

While the EW contribution is smaller than the current discrepancy, contributions from SUSY~\cite{Ellis:1982by} 
with EW scale superpartners may still account for the effect as they may be enhanced by $\tan\beta$, 
the ratio of the vacuum expectation values of the two Higgs doublets.  
Within a global SUSY analysis of EW precision data including $g_\mu - 2$~\cite{Erler:1998ur} 
it has been shown recently that the most plausible scenario is that loop diagrams containing 
a muon-sneutrino, a wino and a charged Higgsino, dominate~\cite{Cho:2011rk}.
This can be achieved already at moderately large $\tan\beta \sim 10$.  
These results are not directly affected by the non-observation of colored superpartners
with masses up to about 1~TeV at the LHC~\cite{delRe:PAVI2011}.

\section{Practical example: gauge extensions}

$Z^\prime$ bosons~\cite{Langacker:2008yv} are among the best motivated kinds of physics beyond the SM.
They easily appear in top-down scenarios like Grand Unified Theories or superstring constructions. 
In fact, it often requires extra assumptions if one wants to {\em avoid\/} an additional $U(1)^\prime$ gauge symmetry 
or decouple the associated $Z^\prime$ from observation. 
This is even more true in bottom-up approaches where $U(1)^\prime$ symmetries are a standard tool 
to alleviate problems in models of dynamical symmetry breaking, supersymmetry, large 
or warped extra dimensions, little Higgs, {\em etc.} 
And as all these models are linked to electroweak symmetry breaking, the $Z^\prime$ mass, $M_{Z^\prime}$, 
should be in the TeV region, providing a rationale why they might be accessible at current or near future experiments. 

$Z^\prime$ discovery would most likely occur as an $s$-channel resonance at a collider, 
but interference with the photon or the standard $Z$ provides leverage also at lower energies. 
Once discovered at a collider, angular distributions may give an indication of its spin to 
discriminate it against states of spin~0 ({\em e.g.\/}, the sneutrino) and spin~2 (like the Kaluza-Klein graviton 
in extra dimension models). The diagnostics of its 
charges would be of utmost importance as they can hint at the underlying principles.

\begin{figure}
\includegraphics[scale=0.241]{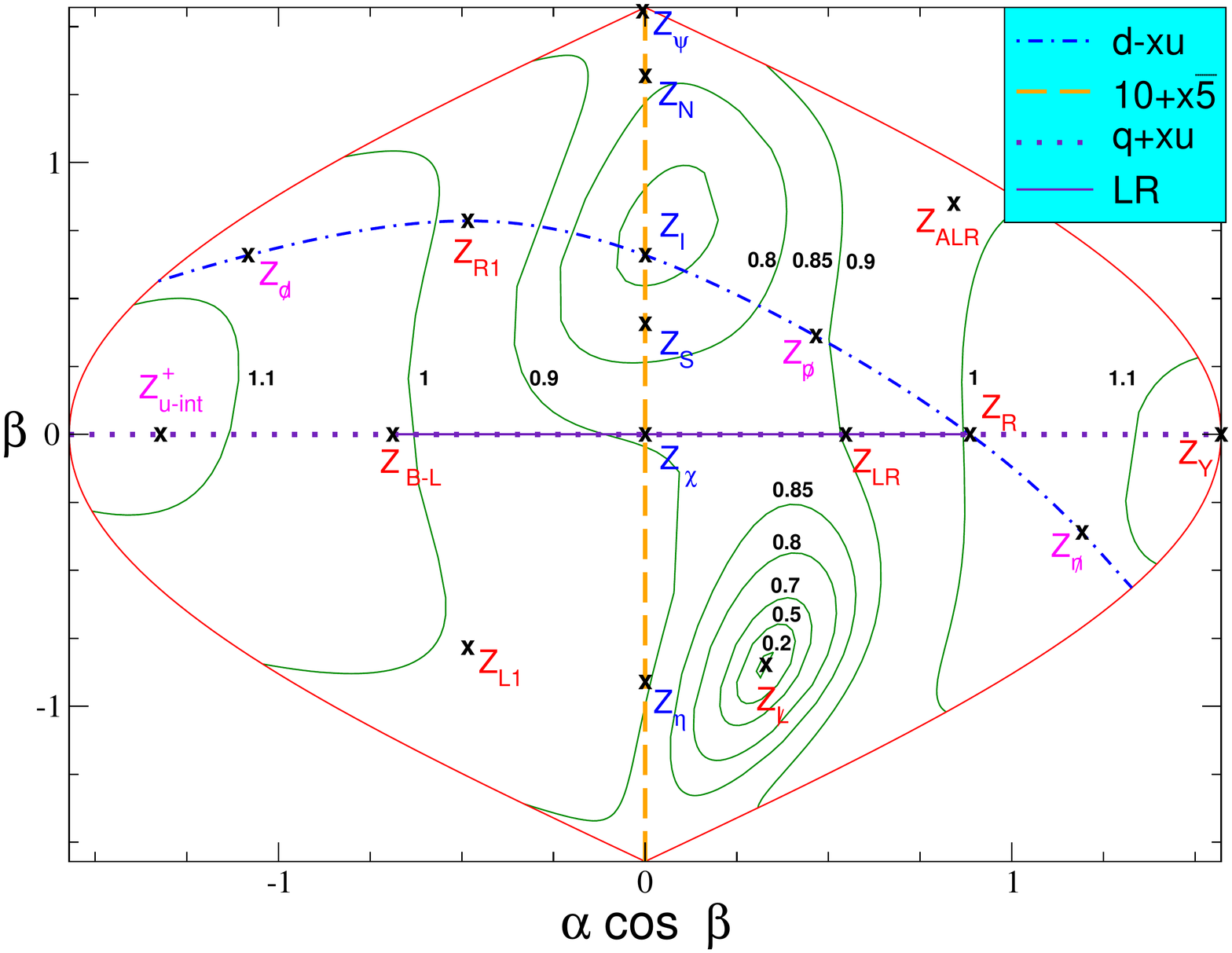}
\includegraphics[scale=0.241]{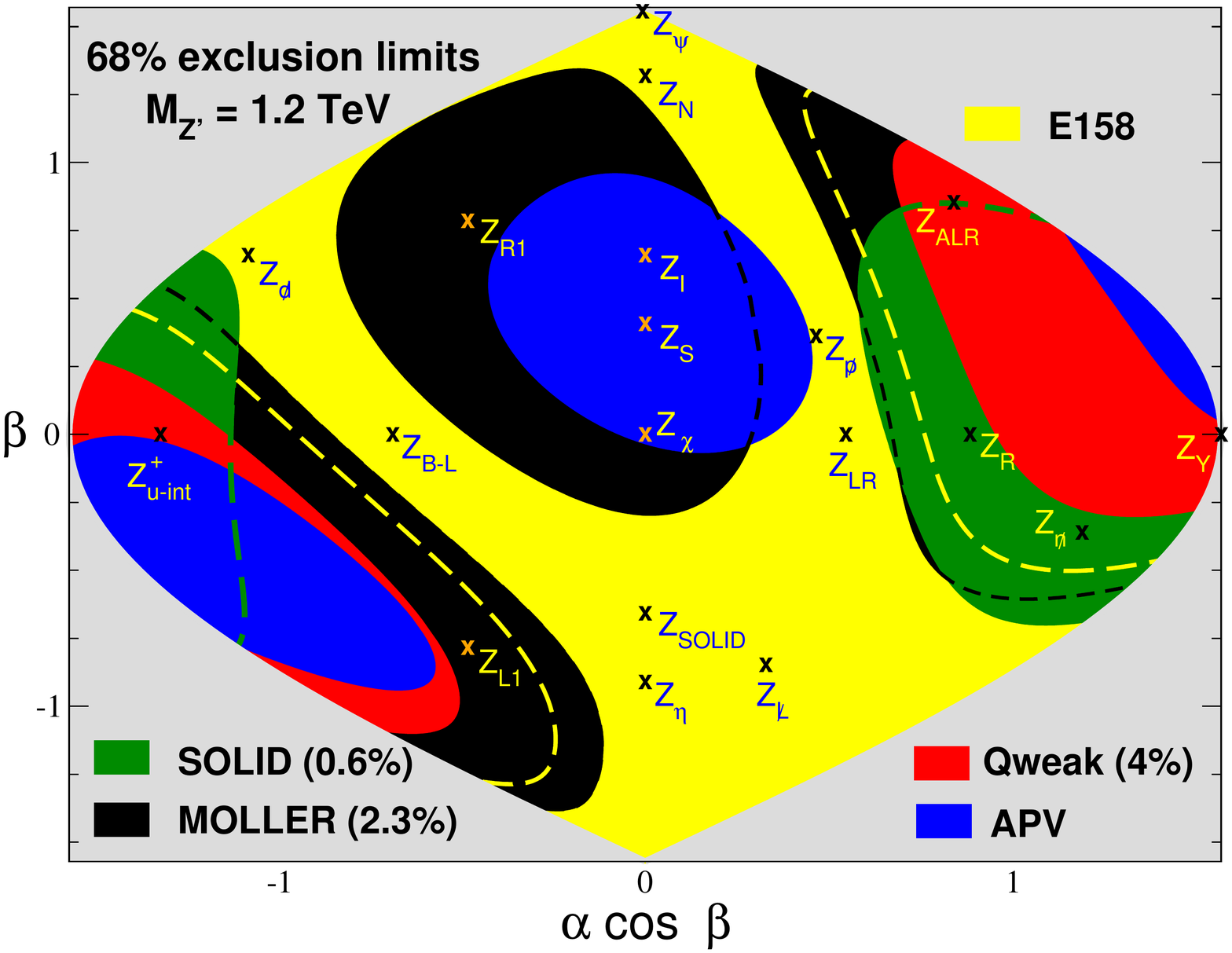}
\caption{Constraints on the $E_6$ parameters $\alpha$ and $\beta$. 
Left: 95\% CL contour lines of $M_{Z^\prime}$ exclusions (in TeV)~\cite{Erler:2011ud} 
exemplified by the CDF di-muon data~\cite{Aaltonen:2008ah}.
Right: 68\% exclusion constraints for $M_{Z^\prime} = 1.2$~TeV from various
actual and hypothetical low energy measurements.}
\label{zprime}
\end{figure}

An interesting class of models is related to $E_6$, a plausible gauge group for unified model building. 
All representations of $E_6$ are free of anomalies so that its $U(1)^\prime$ subgroups 
correspond to $Z^\prime$ candidates.
$Z^\prime$ bosons with the same charges for the SM fermions as in $E_6$ 
also arise within a bottom-up approach~\cite{Erler:2000wu}
when anomaly cancellation is demanded in supersymmetric extensions of the SM 
together with a set of fairly general requirements. 
The breaking chain, $E_6 \to SO(10) \times U(1)_\psi \to SU(5)\times U(1)_\chi \times U(1)_\psi$,
defines a 2-parameter class of models,
$Z^\prime = \cos\alpha \cos\beta Z_\chi + \sin\alpha \cos\beta Z_Y + \sin\beta Z_\psi$,
where $Y$ denotes hypercharge, and
$\alpha \neq 0$ corresponds to the presence to a kinetic mixing term 
$\propto F^{\mu\nu} _YF_{\mu\nu}^\prime$.

\section{Parity violation in electron scattering, atoms and ions}
Figure~\ref{zprime} shows how collider data and low energy constraints from
existing and future polarized electron scattering experiments and atomic parity violation (APV) 
provide complementary information on the parameters introduced in the previous section.
Notice, that the combined data from $e^-$ scattering alone 
may cover the entire parameter space (for future measurements it is assumed that the
central values will coincide with the SM).

\begin{figure}
\includegraphics[width=0.48\textwidth]{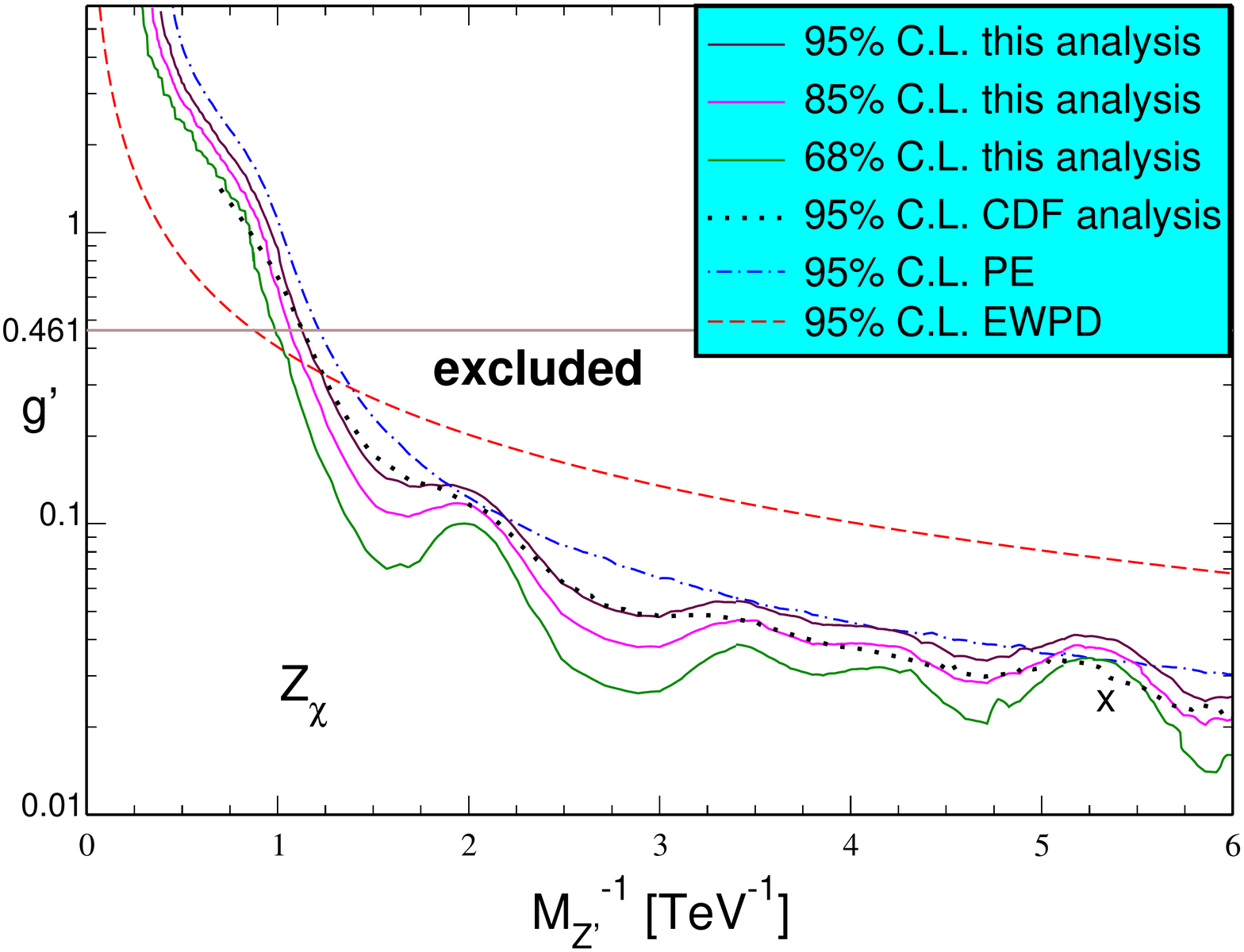}
\includegraphics[width=0.56\textwidth]{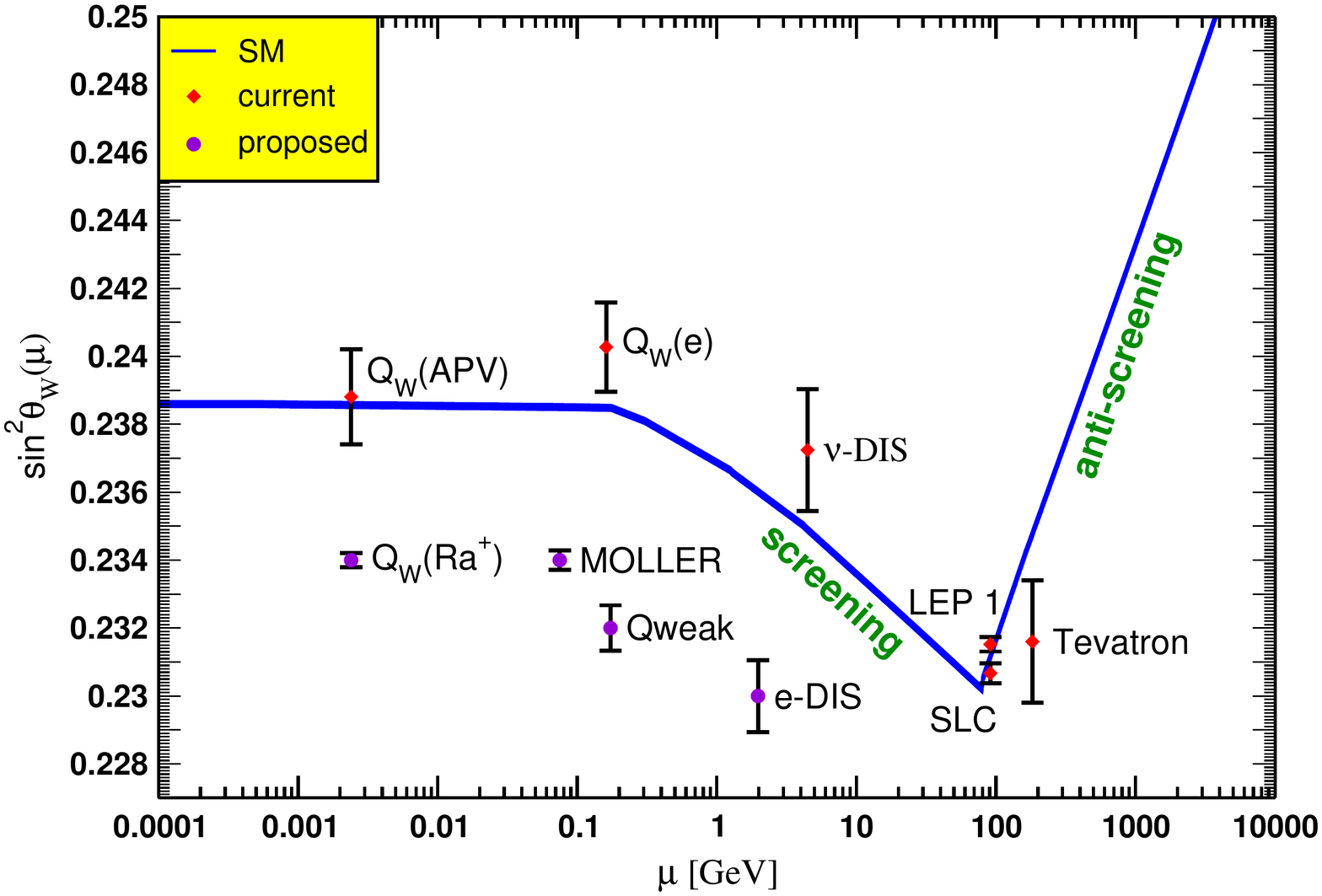}
\caption{Left: EWPD are sensitive to different regions in $Z^\prime$ coupling {\em vs.\/} mass,
as is best seen by comparing with pseudo-experiments (PE).  The wiggles in the lines denoted
"this analysis" (which refers to Ref.~\cite{Erler:2011ud}) and the softer ones from CDF (which is based
on the same data but a different statistical method) are from actual fluctuations in the data.
Right: Various current and future measurements of the running weak mixing angle.
The uncertainty in the prediction is small except possibly in the hadronic transition region roughly
between 0.1 and 2~GeV~\cite{Erler:2004in}.}
\label{escattering}
\end{figure}

Figure~\ref{escattering} illustrates this complementarity in different contexts. 
The left-hand side shows the $Z^\prime$ coupling strength, $g^\prime$, as a function of $M_{Z^\prime}$.  
Collider data produce a stronger $M_{Z^\prime}$ dependance, 
but there is no resemblance of a sharply edged shape~\cite{Erler:2011ud} 
as one might na\"ively expect from di-muon bump hunting. 
The EW precision data (EWPD) result in weaker $M_{Z^\prime}$ dependance
and stronger exclusions at large $g^\prime$.
The right-hand side of Fig.~\ref{escattering} shows the scale dependance of $\sin^2\theta_W$.
Collider experiments measure $Z$ couplings with great precision but are virtually blind to new physics amplitudes.  
They serve to fix the position of the curve. 
The converse is true of electron scattering and APV 
(and also of a possible experiment~\cite{Willmann:PAVI2011} in a single trapped Ra ion).  

It should be stressed that these experiments provide determinations of $\sin^2\theta_W$ that are competitive 
with the $Z$ pole, and that they are also sensitive to supersymmetry, especially when R-parity is 
broken~\cite{RamseyMusolf:2006vr}.

\section{Electric Dipole Moments}
In any relativistic theory permanent EDMs  
(for a review, see Ref.~\cite{Pospelov:2005pr}) of half-integer spin systems
violate time-reversal symmetry~\cite{Weinberg:1995mt}, 
and in any relativistic quantum field theory they violate parity (P) and charge-parity (CP), as well.
Unlike the complex phase in the CKM matrix, CP violation (CPV) due to EDMs is not
related to flavor change. CPV from the CKM matrix is too small 
to produce the baryon asymmetry of the universe which is an important observational
piece of data in support of physics beyond the SM and is --- as far as we know --- 
independent of the hierarchy problem, dark matter and dark energy.  
At the same time, the CKM mechanism is also unable to produce EDMs large enough to
be observable in the foreseeable future. On the other hand, new particles beyond the SM
tend to give rise to new operators generally adding many complex phases and to
EDMs too large to be consistent with current limits (CP problems).

There is a second mechanism for CPV in the SM in form of the so-called QCD $\theta$-term,
\begin{equation}
  {\cal L}_{\bar\theta} = \bar\theta {g_s^2\over 32 \pi^2} G_{\mu\nu}^a \tilde{G}_{\mu\nu}^a, 
\end{equation}
which is a total derivative and also violates P and CP.  
It affects EDMs of hadrons which can be estimated in chiral perturbation theory.
The proton and neutron EDMs, $d_p$ and $d_n$, are dominated by chiral logarithms~\cite{Crewther:1979pi}
while these cancel to leading order for the deuteron. 
The same kind of logarithms enter chromo-electric and gravitational dipole moments,
A recent chiral perturbation theory calculation~\cite{Mereghetti:2010kp} resulted in
$|d_n| \approx |d_p| \approx 2.1 \times 10^{-3}\ \bar\theta$~e~fm.
The experimental limit~\cite{Baker:2006ts},
$|d_n| < 2.9 \times 10^{-13}$~e~fm, then gives the bound, $\bar\theta \lesssim 10^{-10}$.
The unexplained smallness of $\bar\theta$ is known as the strong CP problem.

EDMs strongly constrain models and parameters beyond the SM including those motivated by 
baryogenesis, but is the absence of any non-vanishing EDM already a problem for baryogenesis itself?
This question has been addressed in a simple toy model~\cite{Grojean:2004xa,Huber:2006ri},
\begin{equation}
   {\cal L} = {(H^\dagger H)^3\over \Lambda_{CP}^2} + Z_t (H^\dagger H) \bar{Q}_3 H t,
\end{equation}
with the result that a realistic baryon density $\eta_B$ of ${\cal O}(10^{-10})$ can be achieved
if $ \Lambda_{CP}$ is between 400 and 800~GeV.
Since the next generation of EDM experiments will probe $\Lambda_{CP} \sim 3$~TeV, 
this shows that the simplest scenarios will be excluded if no EDM is seen.

\acknowledgments
It is a pleasure to thank the organizers for the invitation and an enjoyable workshop.
I also like to thank Eduardo Rojas for creating several plots
and also together with Paul Langacker and Shoaib Munir for  collaboration 
on some of the topics presented here.
This work was supported by CONACyT project 82291--F and by PASPA (DGAPA--UNAM).

\end{document}